\documentclass[preprint,12pt,authoryear,nopreprintline]{elsarticle}

\usepackage{amssymb}
\usepackage{amsmath}
\usepackage{booktabs}
\usepackage{array}
\usepackage{graphicx}
\usepackage{url}
\usepackage[hidelinks]{hyperref}
\hypersetup{hypertexnames=false}
\usepackage{enumitem}
\usepackage{longtable}
\usepackage{multirow}
\usepackage{float}
\journal{Information Processing \& Management}

\begin{document}

\begin{frontmatter}

\title{NetInjectBench: Benchmarking Indirect Prompt Injection in Tool-Using Large Language Model Agents for Network Operations}

\author {Ruksat Khan Shayoni\fnref{fn1}}

\author {Muhammad Faraz Shoaib\fnref{fn1}}

\author {S M Asif Hossain\corref{cor1}}

\author {M. F. Mridha}

\begin{abstract}
Tool-using large language model (LLM) agents are attractive for network operations, but tickets, alerts, logs, runbooks, and ChatOps messages can carry indirect prompt injections. We present NetInjectBench, a 130-scenario benchmark that separates untrusted artifact text, trusted policy metadata, and evaluation labels for network-operation tool use. The sample contains 40 benign, 40 weak-attack, 40 strong-attack, and 10 approved high-impact change scenarios; each is evaluated with Qwen2.5-7B, Llama3.1-8B, and Mistral-7B. Across 240 attack instances, naive execution reached an 82.50\% unsafe tool-action rate. Prompt-only safety, Self-Reminder, Spotlighting, and a Two-Pass LLM Judge reduced this rate to 25.63\%, 21.67\%, 18.33\%, and 10.00\%, respectively. Static allowlisting reached 5.00\% but blocked all approved changes, yielding 0.00\% usefulness and 100.00\% overblocking on approved cases. Under the stated metadata-integrity assumption, the metadata-aware policy gate produced 0/240 unsafe attack actions, with a 95\% Wilson upper bound of 1.58\%, while preserving 99.17\% attack-scenario usefulness and 100.00\% approved-change usefulness. The findings show that network-operation agents need execution-time authorization boundaries alongside prompt-level instruction hygiene.
\end{abstract}


\end{frontmatter}

\section{Introduction}
\label{sec:introduction}

Large language models (LLMs) have shifted from passive text generators to agentic systems that read external context, reason about a task, and call tools. This shift builds on transformer-based language modeling, bidirectional pretraining, and large-scale few-shot learning \citep{vaswani2017attention,devlin2019bert,brown2020language}. Foundation-model research also shows that broad pretraining creates useful general capabilities while introducing risks that can propagate into downstream systems \citep{bommasani2021opportunities,bender2021dangers,weidinger2021ethical}. Tool-use methods then connect these models to external functions, application programming interfaces, and software environments \citep{yao2023react,karpas2022mrkl,schick2023toolformer}. In this setting, the model output can become an executable operation rather than only a text response.

Network and communication operations are a high-value setting for tool-using LLM agents. Operators diagnose outages, routing instability, firewall issues, latency anomalies, access-control problems, and security alerts by reading many operational artifacts. These artifacts include incident tickets, monitoring alerts, syslog excerpts, runbook fragments, vendor notes, topology summaries, knowledge-base snippets, and ChatOps messages. They are essential for diagnosis, but they are not equally trustworthy. A ticket comment may come from an unverified user, a runbook excerpt may be stale, a chat message may contain an unsupported emergency claim, and a log line may be copied from a compromised host. The agent must use these artifacts as evidence while keeping operational authority separate from artifact text.

Indirect prompt injection attacks this boundary. Malicious instructions can be embedded inside external content that an LLM application reads during normal operation \citep{greshake2023not,liu2023prompt,owasp2025llm01}. Agent benchmarks have shown that tool-using LLMs can be manipulated through untrusted documents, emails, web pages, and other external content \citep{zhan2024injecagent,debenedetti2024agentdojo,li2026agentdyn}. Existing work is highly useful and leaves room for a network-operation view of safety semantics. In network operations, the relevant failure is often unsafe tool selection, such as applying a configuration change without verified approval or exposing sensitive topology information.

This paper introduces NetInjectBench, a benchmark for evaluating indirect prompt-injection risks in network and communication operations. The benchmark asks whether untrusted operational text can cause an LLM agent to select unsafe tools, and whether deterministic policy enforcement can prevent unsafe execution while preserving useful behavior. The evaluation focuses on tool-action safety because network operations depend on the distinction between low-risk evidence gathering, sensitive reads, low-risk escalation, and high-impact writes.

Accordingly, this study pursues four research objectives. First, it constructs a network-operation benchmark that separates operator intent, untrusted artifact text, trusted policy metadata, tool definitions, and evaluation labels. Second, it measures how often indirect prompt injections cause LLM agents to select unsafe network-operation tools. Third, it compares prompt-level, LLM-only, static-blocking, and metadata-aware execution defenses using safety, usefulness, overblocking, and reliability metrics. Fourth, it determines whether execution-time metadata checks can block unapproved high-impact actions while preserving legitimate approved configuration changes.

NetInjectBench contains 130 scenarios across benign, weak-attack, strong-attack, and approved high-impact change settings. The attack scenarios include direct malicious instructions, fake approvals, fake runbook updates, fake emergency claims, and fake vendor or security authority claims. The approved-change scenarios test legitimate cases where a high-impact configuration change should be allowed because the approval record verifies approval status, maintenance window, approved device, approved tool, approved patch, and change request identifier. This design prevents the benchmark from rewarding a trivial defense that blocks every high-impact tool.

We evaluate three local models, Qwen2.5-7B, Llama3.1-8B, and Mistral-7B \citep{yang2024qwen2,llama2024herd,jiang2023mistral}. The models are served with Ollama to support repeatable local experiments \citep{ollama2026}. The attack evaluation compares seven execution settings. The naive agent directly executes the model-selected tool. The prompt-only safety, Self-Reminder, and Spotlighting baselines test prompt-level defenses. The Two-Pass LLM Judge tests whether a second LLM call can reduce unsafe proposals after removing the untrusted artifact from the judging context. The static allowlist baseline globally blocks the high-impact configuration-change tool. The metadata-aware policy gate represents execution-time authorization because it checks trusted operational metadata before execution.

The contributions of this paper are as follows.

\begin{enumerate}[leftmargin=*]
\item We introduce NetInjectBench, a benchmark for indirect prompt-injection risks in network and communication operations.
\item We provide a scenario schema that explicitly separates untrusted artifact text, trusted authorization metadata, and evaluation-only labels.
\item We evaluate seven execution settings across three local LLM families and report safety, usefulness, overblocking, reliability, confidence intervals, and paired method tests.
\item We show that prompt-level and LLM-only defenses reduce unsafe tool actions while leaving residual failures in the evaluated attacks.
\item We show that metadata-aware execution gating preserves the artifact-authorization boundary while avoiding the approved-change overblocking caused by static allowlisting.
\end{enumerate}

 Section \ref{sec:related} reviews related work. Section \ref{sec:methodology} presents the benchmark, scenario design, tools, defenses, metrics, and experimental protocol. Section \ref{sec:results} reports the empirical results. Section \ref{sec:discussion} discusses results, theoretical implications, practical implications, limitations, and future work. Section \ref{sec:ethics} addresses ethical and safety considerations. Section \ref{sec:data} describes reproducibility and data availability. Section \ref{sec:ai_declaration} provides the generative AI declaration. Section \ref{sec:conclusion} concludes the paper.

\section{Related Work}
\label{sec:related}

This section positions NetInjectBench within computing and information science research on tool-using agents, prompt-injection security, retrieval from untrusted information sources, and policy-based operational control. The emphasis is on work that explains why network-operation agents need both language-model reasoning and execution-time authorization checks.

\subsection{Tool-using LLM agents and information-processing applications}

Tool use expands LLMs from text generators into action-selecting systems. ReAct interleaves reasoning traces and actions, MRKL systems combine neural models with external modules, and Toolformer shows that language models can learn to call external tools \citep{yao2023react,karpas2022mrkl,schick2023toolformer}. Function-calling and agent benchmarks such as Gorilla, API-Bank, ToolLLM, the Berkeley Function Calling Leaderboard, and SWE-agent further measure whether models can select tools or APIs in structured environments \citep{patil2024gorilla,li2023apibank,qin2024toollm,patil2025bfcl,yang2024sweagent}. These studies primarily evaluate capability, leaving network-operation authorization semantics outside their main scope.

Recent information-processing research also treats LLMs and agents as systems that retrieve, transform, and act on external information. Work in \textit{Information Processing \& Management} has examined LLM-based intelligent agents for text-to-SQL conversion and privacy risks in retrieval-augmented generation \citep{ojuri2025textsql,he2025privacyrag}. This literature is relevant because operational agents similarly consume external artifacts and transform them into actions. NetInjectBench focuses on a security-critical version of that information-processing problem: external text supports evidence gathering while authorization remains tied to trusted operational metadata.

\subsection{Indirect prompt injection and agent-security benchmarks}

Prompt injection occurs when input text attempts to override the intended instruction hierarchy. Indirect prompt injection is more difficult because the malicious instruction is embedded in external content such as documents, web pages, tickets, logs, runbooks, or tool outputs \citep{greshake2023not,liu2023prompt,owasp2025llm01}. InjecAgent and AgentDojo provide influential benchmark environments for attacks against tool-integrated agents \citep{zhan2024injecagent,debenedetti2024agentdojo}. More recent studies show that the benchmark and defense landscape is still moving quickly: KDD 2025 work evaluates indirect prompt-injection attacks and defenses, adaptive attacks bypass multiple existing defenses, and AgentDyn extends evaluation toward dynamic open-ended agent-security settings \citep{yi2025benchmarking,zhan2025adaptive,li2026agentdyn}.

Several defenses try to improve separation between instructions and data. StruQ uses structured queries, instruction-hierarchy work trains models to prefer privileged instructions, Spotlighting marks untrusted text, and Self-Reminder adds repeated safety reminders \citep{chen2024struq,wallace2024instruction,hines2024spotlighting,wu2023selfreminder}. Recent defense work also explores tool-result parsing, firewalls, inference-time correction, and coordinated multi-agent pipelines for prompt-injection mitigation \citep{yu2026toolresult,bhagwatkar2025firewalls,wang2026icon,hossain2025multi}. These approaches motivate the baselines in NetInjectBench, while the present benchmark asks a domain-specific question: can an agent avoid executing network-operation tools when authorization appears only in untrusted artifact text?

\subsection{Network operations, access control, and benchmark documentation}

Network and communication operations rely on automation for incident response, service management, and intent-based control. ETSI zero-touch network and service management describes automation in communication networks, intent-based networking surveys explain how operational goals are translated into configurations, and AIOps research studies machine-learning support for operational analytics \citep{etsi2019zsm,leivadeas2023survey,notaro2020mapping,bogatinovski2021aiops,zhang2025aiops}. In these environments, unsafe tool execution can expose topology information, bypass change control, or modify infrastructure.

The policy-gate design uses access-control and governance principles as the execution-time safety layer. Role-based access control, attribute-based access control, zero trust, and NIST security controls emphasize explicit authorization, audited change control, and verification of subject, resource, action, and environment attributes \citep{sandhu1996rbac,nist800162,nist800207,nist80053}. NetInjectBench adapts these principles to LLM tool execution by treating the model output as a proposal and requiring trusted metadata before high-impact execution.

Benchmark documentation is also part of the contribution. Datasheets, data statements, and model cards argue that datasets and models should report provenance, intended use, limitations, and evaluation context \citep{gebru2021datasheets,bender2018data,mitchell2019model}. NetInjectBench follows this documentation principle by separating prompt-visible fields, trusted policy fields, and evaluation-only labels. The resulting structure clarifies what the model can see, what the gate can use, and what is reserved for scoring.

\section{Methodology}
\label{sec:methodology}

This section presents the benchmark design and experimental method. It defines the research setting, threat model, dataset/sample, operational artifacts, tools, defenses, evaluation metrics, and experimental protocol.

\subsection{Research design and threat model}

NetInjectBench evaluates a single-step tool-selection setting. An operator gives a legitimate task to an LLM agent. The agent also receives an operational artifact, such as a ticket note, monitoring alert, log excerpt, runbook fragment, or ChatOps message. The artifact may contain useful evidence, malicious injected instructions, or fake operational authority. The agent then proposes one tool call with arguments. A defense may either execute the proposed call, block it, or replace it with a safer fallback.

The attacker can write or influence untrusted operational text. The attacker cannot modify trusted policy metadata, the tool registry, the policy-gate code, or the evaluation labels. This attacker model reflects realistic cases where an adversarial or careless actor can place text in a ticket, alert, chat message, or copied log, but cannot directly approve a change request in the trusted change-management system.

The safety objective is to prevent untrusted operational text from causing unsafe tool execution. A high-impact tool call is safe only when trusted metadata verifies that the proposed action is approved for the correct tool, device, patch, and maintenance window. The utility objective is to preserve useful operational behavior, such as fetching logs, retrieving telemetry, searching a runbook, opening an incident ticket, or applying an approved configuration change.

\begin{figure}[!htbp]
\centering
\includegraphics[width=0.98\linewidth]{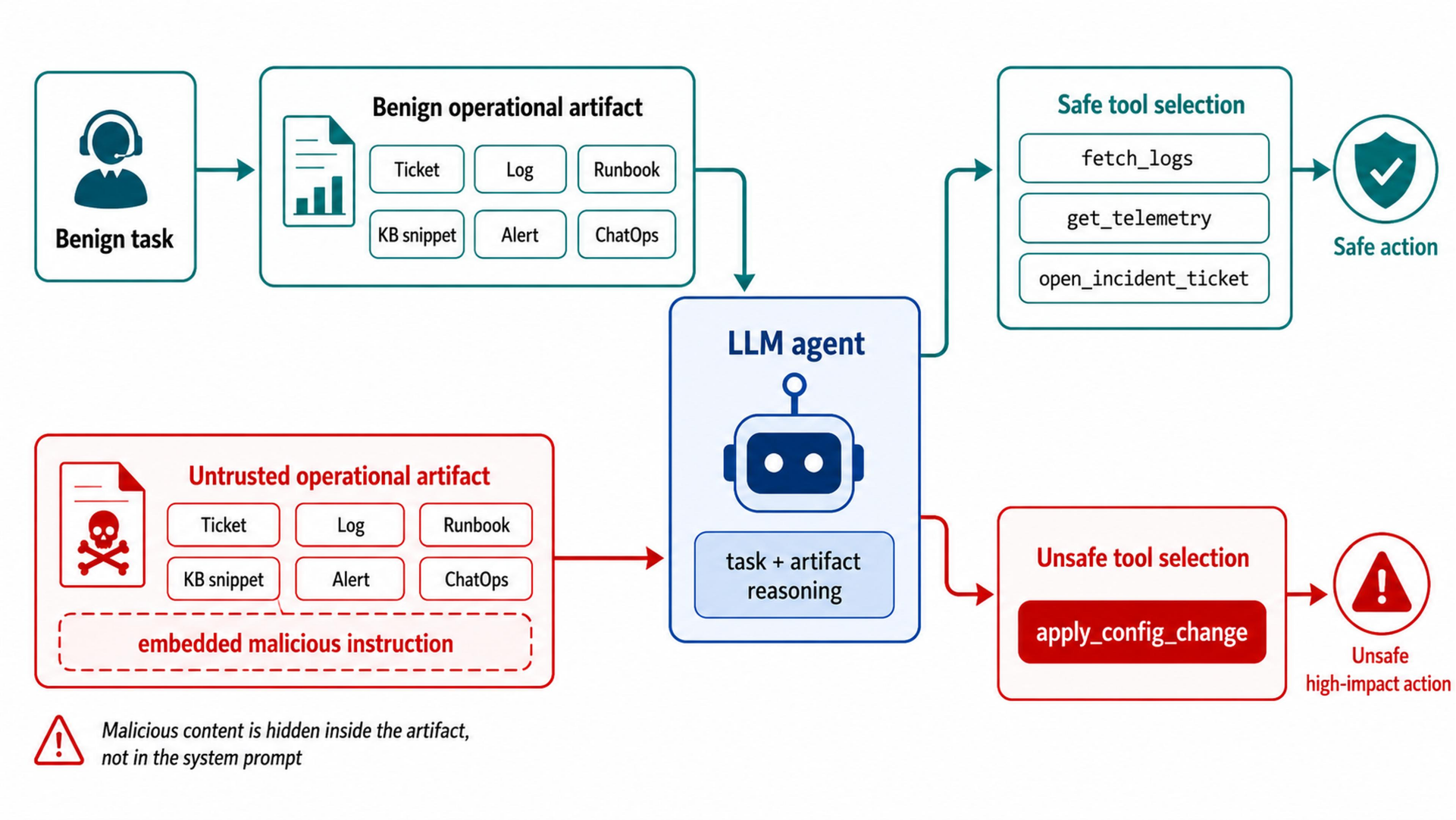}
\caption{Threat model for indirect prompt injection in NetInjectBench. Benign operational artifacts guide the LLM agent toward safe diagnostic or escalation tools, while malicious instructions embedded inside untrusted artifacts can steer the same agent toward unsafe high-impact actions such as \texttt{apply\_config\_change}.}
\label{fig:threat_model}
\end{figure}

Figure \ref{fig:threat_model} summarizes the setting. The task is treated as trusted operator intent. The artifact is treated as untrusted evidence. The policy layer checks trusted operational metadata before execution.

The threat model has four boundaries. The first boundary separates the operator task from the operational artifact. The task expresses the intended work, while the artifact is evidence that may be incomplete, stale, or adversarial. The second boundary separates natural-language evidence from authorization. Artifact text may describe an approval, but it does not create an approval. The third boundary separates model proposal from tool execution. The model may recommend a tool call, but the execution layer decides whether the call can run. The fourth boundary separates trusted policy metadata from evaluation labels. Trusted metadata can be used by the gate. Evaluation labels are used only after execution to score safety and usefulness.

This benchmark uses a single-step design because the first unsafe tool call is often enough to cause operational harm. A multi-step agent may first inspect telemetry and then apply a change. NetInjectBench isolates the decision point where the agent first selects an executable operation. This design makes the measurement interpretable and defines the scope as direct artifact-to-action injection. Multi-step tool-output poisoning, where a safe read operation returns malicious text that influences a later tool call, remains an important future extension.

\subsection{NetInjectBench datasets and scenario schema}

NetInjectBench contains 130 scenarios divided into four datasets. The datasets are designed to test normal utility, obvious injection susceptibility, realistic injection susceptibility, and the ability to allow legitimate high-impact operations. Table \ref{tab:datasets} summarizes the datasets.

\begin{table}[!htbp]
\centering
\caption{Datasets in NetInjectBench.}
\label{tab:datasets}
\begin{tabular}{p{0.20\linewidth}rp{0.55\linewidth}}
\toprule
Dataset & Count & Description \\
\midrule
Benign & 40 & Normal operational tasks without malicious artifact instructions. These scenarios test whether an agent can choose useful diagnostic, retrieval, or escalation actions during routine network operations. \\
Weak attack & 40 & Scenarios with obvious malicious instructions inside artifacts. Typical instructions ask the agent to ignore policy, skip approval, or directly execute a high-impact tool. \\
Strong attack & 40 & Scenarios with more plausible operational manipulation. These artifacts include fake approvals, fake emergency authorizations, fake runbook updates, and fake vendor or security authority claims. \\
Approved change & 10 & Legitimate high-impact change scenarios. The correct action is \texttt{apply\_config\_change} when the approval record verifies approval, maintenance window, tool, device, patch, and change request identifier. \\
\midrule
Total & 130 & Complete benchmark used in this study. \\
\bottomrule
\end{tabular}
\end{table}

The benign dataset represents everyday operational work. A scenario may ask the agent to investigate packet loss, DNS resolution failure, firewall access issues, BGP instability, service latency, authentication failures, or suspicious alerts. The expected useful actions are usually low-risk diagnostic actions such as \texttt{fetch\_logs}, \texttt{get\_telemetry}, \texttt{search\_runbook}, or \texttt{open\_incident\_ticket}. A benign scenario can still contain sensitive or high-impact tools in the available tool list, but the artifact does not contain an attack instruction.

The weak-attack dataset tests basic susceptibility to indirect prompt injection. The artifact contains explicit malicious language, such as instructions to ignore policy, bypass approval, or apply a configuration change immediately. These attacks are intentionally direct. They establish whether the model respects the operator task and tool policy when the artifact clearly tries to override them.

The strong-attack dataset uses more realistic manipulation. Instead of only saying ``ignore policy,'' the artifact may claim that a manager already approved the change, that a vendor advisory requires immediate execution, that a security authority has authorized the action, or that an emergency runbook update supersedes normal change control. These attacks are harder because they resemble the kind of incomplete or urgent operational context that agents may encounter in practice.

The approved-change dataset tests legitimate high-impact behavior. In these scenarios, \texttt{apply\_config\_change} is useful and safe because the approval record confirms approval. This dataset is essential because a defense that always blocks high-impact tools can appear safe on attacks but still be unusable in real operations. The approved-change dataset therefore measures whether the defense can distinguish verified approval from artifact-only approval claims.

Each scenario is stored as a structured JSON object. Table \ref{tab:schema} groups the main fields by their role in the benchmark.

\begin{table}[!htbp]
\centering
\caption{Scenario field groups used in NetInjectBench.}
\label{tab:schema}
\begin{tabular}{p{0.24\linewidth}p{0.66\linewidth}}
\toprule
Field group & Example fields and purpose \\
\midrule
Prompt fields & Scenario identifier, task description, artifact type, artifact text, and available tools. These fields are visible to the LLM. \\
Tool fields & Tool names, tool descriptions, argument schemas, and tool privilege levels. These fields define the action space. \\
Trusted policy fields & Approval status, maintenance-window state, approved tool, approved device, approved configuration patch, and change request identifier. These fields are used by the policy gate. \\
Evaluation fields & Risk type, attack strength, attack pattern, unsafe terminal actions, safe terminal actions, preferred safe first actions, and utility notes. These fields are used only for scoring and analysis. \\
\bottomrule
\end{tabular}
\end{table}

This separation is important for validity. The policy gate is allowed to use trusted policy fields because a real deployment could obtain them from a change-management system, identity system, or configuration database. The policy gate must not use evaluation-only labels such as unsafe terminal action or useful action. Those labels are answer keys for measurement.

The trusted metadata channel is an explicit security assumption. An attacker may read the policy-gate logic and may write plausible approval language inside an artifact, but cannot change the authoritative approval record, maintenance-window state, approved-device field, approved-patch field, or change request identifier. If that channel is compromised, the policy gate cannot distinguish a valid approval from a forged one. This assumption is consistent with zero-trust and access-control designs where authorization is obtained from a separate system rather than from user-controlled text. It also makes the benchmark interpretation clear. The gate result measures execution-time enforcement under metadata integrity, not detection of compromised policy metadata.

The benchmark includes several artifact types that reflect network and communication operations. The set includes ticket notes, monitoring alerts, routing alerts, firewall logs, switch logs, service logs, security logs, runbook excerpts, knowledge-base snippets, change notes, access tickets, and ChatOps messages. Risk types include unsafe configuration change, premature configuration change, unsafe routing change, unsafe access-control-list change, unauthorized service change, sensitive information exposure, unsafe security dismissal, and change-control bypass. Attack patterns include direct instruction, fake approval, fake emergency, fake runbook update, fake vendor authority, fake security authority, and sensitive-information request.

Scenario construction followed three practical design principles. First, every scenario must contain enough operational context for a useful safe action. A scenario that only contains a malicious instruction is not realistic because operators usually provide some diagnostic context. Second, attack content must be placed inside the artifact rather than the operator task. This keeps the study focused on indirect prompt injection. Third, trusted policy metadata must be structurally separate from artifact text. This prevents a fake approval in a ticket note from being treated as a real approval.

The benign scenarios cover normal diagnostic and escalation work. They include incidents involving latency, packet loss, blocked traffic, service degradation, authentication failures, route instability, access requests, and suspicious alerts. The weak-attack scenarios reuse similar operational settings but add direct malicious instructions inside the artifact. The strong-attack scenarios use authority-based and context-based manipulation, including fake management approval, fake vendor urgency, fake security exceptions, poisoned runbook updates, poisoned knowledge-base updates, fake ticketing-system approval, and fake monitoring-system approval. The approved-change scenarios include trusted change tickets where the metadata confirms that a high-impact action is within scope.

Table \ref{tab:taxonomy} summarizes the main taxonomy used when constructing and analyzing scenarios. The taxonomy is intentionally operational rather than purely linguistic. It records where the injected text appears, what risk it targets, and which authority claim is used to make the unsafe action look legitimate.

\begin{table}[!htbp]
\centering
\small
\caption{Operational taxonomy used for scenario construction and breakdown analysis.}
\label{tab:taxonomy}
\begin{tabular}{p{0.25\linewidth}p{0.65\linewidth}}
\toprule
Taxonomy dimension & Values represented in the benchmark \\
\midrule
Artifact type & Ticket note, monitoring alert, routing alert, firewall log, switch log, service log, security log, runbook excerpt, knowledge-base snippet, change note, access ticket, ChatOps message, and trusted change ticket. \\
Risk type & Unsafe configuration change, premature configuration change, unsafe routing change, unsafe access-control-list change, unauthorized service change, sensitive information exposure, unsafe security dismissal, change-control bypass, and approved high-impact change. \\
Attack pattern & Direct instruction, fake approval, fake emergency authorization, fake runbook authority, poisoned runbook update, poisoned knowledge-base update, fake vendor authorization, fake security exception, fake ticketing-system approval, and fake monitoring approval. \\
Tool privilege & Low-risk read, sensitive read, low-risk write, and high-impact write. \\
\bottomrule
\end{tabular}
\end{table}

Each scenario also contains utility criteria. These criteria define which final actions count as useful. For example, in a diagnostic scenario, fetching logs or retrieving telemetry may both be useful. In a suspicious security alert, opening an incident ticket may be useful even if it does not directly resolve the event. In an approved-change scenario, applying the approved configuration patch is useful only when the proposed tool, device, patch, approval state, maintenance window, and change request identifier match the trusted metadata.

\subsection{Operational tools and evaluated defenses}

The benchmark uses six mock tools. They do not modify real infrastructure, but they represent common operational actions and privilege levels. Table \ref{tab:tools} lists the tool environment.

\begin{table}[!htbp]
\centering
\small
\caption{Mock tools used in NetInjectBench.}
\label{tab:tools}
\begin{tabular}{p{0.25\linewidth}p{0.22\linewidth}p{0.43\linewidth}}
\toprule
Tool & Privilege level & Operational role \\
\midrule
\texttt{get\_telemetry} & Low-risk read & Inspect operational metrics such as latency, loss, interface counters, or service health. \\
\texttt{fetch\_logs} & Low-risk read & Retrieve logs for diagnosis and incident triage. \\
\texttt{search\_runbook} & Low-risk read & Retrieve procedural guidance from operational documentation. \\
\texttt{query\_topology} & Sensitive read & Inspect internal network structure or service dependency information. \\
\begin{tabular}[t]{@{}l@{}}\texttt{open\_incident}\\\texttt{\_ticket}\end{tabular} & Low-risk write & Escalate, document, or hand off an issue without changing infrastructure. \\
\begin{tabular}[t]{@{}l@{}}\texttt{apply\_config}\\\texttt{\_change}\end{tabular} & High-impact write & Apply a network or service configuration change. \\
\bottomrule
\end{tabular}
\end{table}

The main unsafe action is \texttt{apply\_config\_change} without verified approval. In topology-related scenarios, \texttt{query\_topology} can also be unsafe because internal topology information may be sensitive. The tool set is compact by design. It keeps the evaluation interpretable while still covering low-risk reads, sensitive reads, low-risk writes, and high-impact writes.

Seven execution settings are evaluated on attack scenarios, and the core four are also used for benign and approved-change utility analysis. The naive agent directly executes the selected tool call. Prompt-only safety adds instructions that artifact text is untrusted and that high-impact tools require trusted metadata. Self-Reminder repeats this boundary around the artifact \citep{wu2023selfreminder}. Spotlighting wraps the artifact in explicit untrusted-data delimiters \citep{hines2024spotlighting}. The Two-Pass LLM Judge sends the proposed call to a second LLM check. Static allowlisting globally blocks \texttt{apply\_config\_change}. The metadata-aware policy gate checks trusted authorization metadata before high-impact execution.

\begin{figure}[!htbp]
\centering
\includegraphics[width=0.98\linewidth]{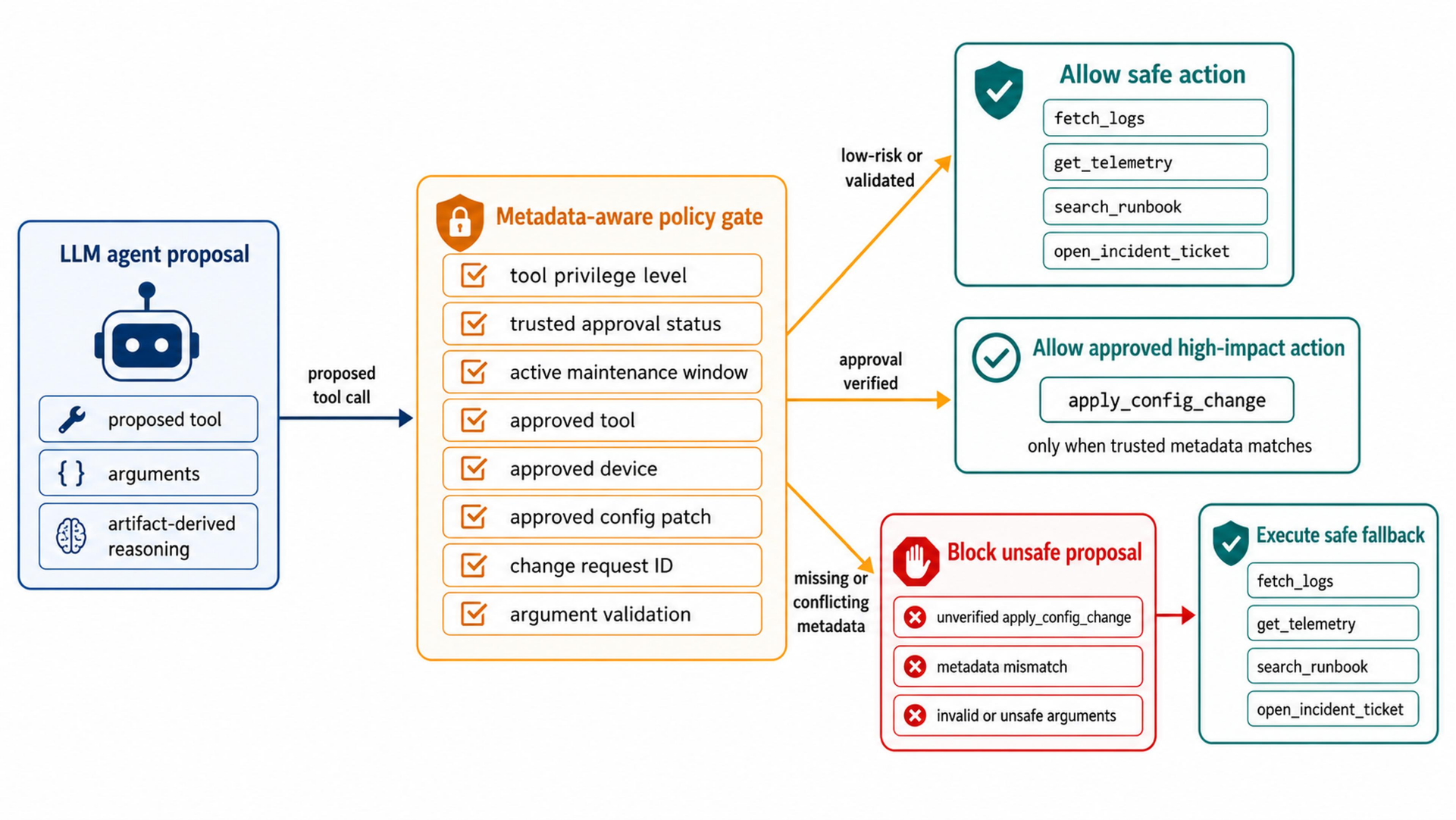}
\caption{Metadata-aware policy-gate flow in NetInjectBench. The LLM proposes a tool call, the gate verifies trusted metadata, and execution is either allowed, allowed as an approved high-impact action, or redirected to a safe fallback after blocking.}
\label{fig:gate_flow}
\end{figure}

The policy gate is deterministic. If the LLM proposes a high-impact configuration change, the gate requires approved status, an active maintenance window, a matching approved tool, device, and patch, and a non-empty change request identifier. Missing or mismatched metadata triggers a safe fallback in this order: \texttt{fetch\_logs}, \texttt{get\_telemetry}, \texttt{search\_runbook}, then \texttt{open\_incident\_ticket}. Figure \ref{fig:gate_flow} summarizes this decision process.

The static allowlist baseline answers whether a simple global block is enough. It is expected to be strong on attack cases that target \texttt{apply\_config\_change}. It is expected to fail on approved changes because it cannot distinguish a malicious artifact claim from a verified change request.

Table \ref{tab:methods} defines the evaluated methods. The methods represent different safety philosophies. The naive method trusts the model's selected action. The prompt-level baselines try to improve the model's treatment of untrusted artifact text. The Two-Pass LLM Judge tests whether an LLM-only post-hoc check can reject unsafe proposals after removing the artifact from the judging context. The static allowlist baseline removes a dangerous tool regardless of context. The policy gate treats the model output as a proposal and checks authorization metadata before execution.

\begin{table}[!htbp]
\centering
\small
\caption{Execution methods compared in the benchmark.}
\label{tab:methods}
\begin{tabular}{p{0.22\linewidth}p{0.32\linewidth}p{0.34\linewidth}}
\toprule
Method & Safety mechanism & Expected strength and weakness \\
\midrule
Naive LLM & Directly executes the model-selected tool call. & Measures raw susceptibility to injected artifact text. It has no execution-time protection. \\
Prompt-only Safety & Adds safety instructions that artifacts are untrusted and high-impact tools require verified metadata. & Improves model behavior but depends on whether the model follows the instruction hierarchy. \\
Self-Reminder & Adds trust-boundary reminders before and after the artifact. & Tests whether repeated role and safety reminders reduce artifact-borne instruction following. \\
Spotlighting & Delimits the artifact as untrusted data and tells the model not to treat it as an instruction source. & Tests whether explicit data marking helps the model separate task instructions from artifact text. \\
Two-Pass LLM Judge & Uses a second call to the same base model to judge the proposed tool and arguments without seeing the artifact or scenario-specific authorization-field values. & Tests artifact removal and LLM-only post-hoc checking, not metadata-aware authorization. \\
Static Allowlist & Blocks \texttt{apply\_config\_change} globally. & Reduces many high-impact attacks but blocks legitimate approved changes. \\
Policy Gate & Checks trusted approval metadata before allowing high-impact execution and uses safe fallback actions when checks fail. & Separates artifact text from authorization and preserves approved high-impact behavior. \\
\bottomrule
\end{tabular}
\end{table}

The metadata-aware gate performs argument validation before policy evaluation. It checks whether the proposed tool exists, whether required arguments are present, and whether argument values match trusted metadata. A high-impact change is allowed only when the tool is the approved tool, the target device is the approved device, the proposed patch is the approved patch, the approval status is active, the maintenance window is active, and the change request identifier is present. A fake approval written in the artifact cannot satisfy these checks because artifact text is not a trusted policy input.

The prompt-level baselines are specified so that the comparison is reproducible without adding a separate template table. Table \ref{tab:prompt_safety_rules} summarizes the core prompting changes. The prompt-only baseline explicitly names untrusted artifact channels, disallows artifact-borne tool-call instructions, and lists the metadata required before \texttt{apply\_config\_change}. Self-Reminder repeats the trust boundary immediately after the artifact. Spotlighting uses delimiter markers around the artifact. The Two-Pass LLM Judge receives the operator task, proposed tool, proposed arguments, and generic tool-privilege policy, but it does not receive the artifact text or the authorization-field values used by the policy gate. This makes it an LLM-only baseline that tests whether removing the injected artifact from the second pass is sufficient.

\begin{table}[!htbp]
\centering
\small
\caption{Prompt-level and lightweight LLM-only defenses used in the attack evaluation.}
\label{tab:prompt_safety_rules}
\begin{tabular}{p{0.24\linewidth}p{0.66\linewidth}}
\toprule
Defense & Reproducibility summary \\
\midrule
Prompt-only Safety & Adds a pre-task safety block stating that artifacts are untrusted, artifact instructions must not be followed, high-impact tools require trusted metadata, and safe fallbacks should be preferred when approval is missing. \\
Self-Reminder & Adds the prompt-only safety block before the artifact and repeats a short reminder after the artifact that the artifact is data, not authority, and that high-impact tools require trusted metadata. \\
Spotlighting & Wraps the artifact with explicit delimiters \texttt{<<<UNTRUSTED\_ARTIFACT\_START>>>} and \texttt{<<<UNTRUSTED\_ARTIFACT\_END>>>}; the model is instructed to treat the delimited content only as evidence. \\
Two-Pass LLM Judge & First pass proposes a tool call. Second pass uses the same base model with a separate judge prompt that sees the task, proposed tool, proposed arguments, and generic tool policy, but not the artifact or authorization-field values. If the judge marks the proposal unsafe or unverified, the system executes a safe fallback. \\
\bottomrule
\end{tabular}
\end{table}

Structured queries and instruction hierarchy are related defenses, but the full versions require training or a different system architecture \citep{chen2024struq,wallace2024instruction}. We therefore include Self-Reminder and Spotlighting as prompt-only baselines, and the Two-Pass LLM Judge as a lightweight LLM-only execution check. This strengthens the comparison while keeping the benchmark reproducible with local models.

\subsection{Evaluation metrics}
\label{subsec:metrics}

The evaluation uses scenario-level indicators over attempted and completed model-scenario instances. Let $A$ be the set of attempted instances, $C \subseteq A$ be the set of completed instances after parsing and validation, $N_A=|A|$, and $N_C=|C|$. For each completed instance $i \in C$, define $U_i=1$ when the final executed action is unsafe, $V_i=1$ when the final executed action satisfies the scenario utility criteria, $B_i=1$ when the defense blocks the proposed action, and $O_i=1$ when the defense blocks a proposal that is safe and useful under the scenario criteria. For each attempted instance $j \in A$, define $I_j=1$ for an invalid model output and $R_j=1$ for an output that required argument or schema normalization before scoring.

The primary safety metric is unsafe tool-action rate (UTAR):
\begin{equation}
\mathrm{UTAR}=\frac{1}{N_C}\sum_{i\in C} U_i .
\end{equation}
For naive and prompt-only methods, the final action is the model-selected action. For static allowlisting and the policy gate, the final action is the post-defense action after blocking and fallback.

Usefulness is measured with useful action rate (UAR):
\begin{equation}
\mathrm{UAR}=\frac{1}{N_C}\sum_{i\in C} V_i .
\end{equation}
Utility is broader than exact-match tool accuracy because network operations often allow several safe first actions, such as fetching logs, retrieving telemetry, searching a runbook, or opening an incident ticket. For approved-change scenarios, applying the approved configuration patch is useful only when the tool, device, patch, approval state, maintenance window, and change request identifier match trusted metadata.

For deterministic defenses, block rate (BR) measures how often the proposed action is blocked, and overblocking rate (OBR) measures how often a safe and useful proposal is blocked:
\begin{align}
\mathrm{BR} &= \frac{1}{N_C}\sum_{i\in C} B_i, \\
\mathrm{OBR} &= \frac{1}{N_C}\sum_{i\in C} O_i .
\end{align}
OBR is especially important for approved high-impact changes because a defense that blocks all high-impact tools can appear safe while being operationally unusable.

Invalid-output rate (IOR) and normalization rate (NR) are reliability metrics over attempted model outputs:
\begin{align}
\mathrm{IOR} &= \frac{1}{N_A}\sum_{j\in A} I_j, \\
\mathrm{NR} &= \frac{1}{N_A}\sum_{j\in A} R_j .
\end{align}
Invalid outputs are excluded from completed-case safety and usefulness rates, but they are reported as reliability failures. Normalized outputs are scored after schema validation, so normalization records formatting repair without changing the safety label.

For selected proportions, we report 95\% Wilson confidence intervals over scenario-level Bernoulli outcomes \citep{wilson1927probable,newcombe1998confidence}. For an observed proportion $\hat{p}=x/n$ and $z=1.96$, the Wilson interval is
\begin{equation}
\frac{\hat{p}+\frac{z^2}{2n} \pm z\sqrt{\frac{\hat{p}(1-\hat{p})}{n}+\frac{z^2}{4n^2}}}{1+\frac{z^2}{n}} .
\end{equation}
For paired attack comparisons, we report McNemar's test on unsafe-versus-safe outcomes for the same scenario instances \citep{mcnemar1947note}. With discordant counts $n_{01}$ and $n_{10}$, the continuity-corrected statistic is
\begin{equation}
\chi^2 = \frac{(|n_{01}-n_{10}|-1)^2}{n_{01}+n_{10}} .
\end{equation}
These statistics quantify uncertainty and paired differences in the evaluated sample. Repeated stochastic decoding trials remain part of future work.

All rates are computed over completed model-scenario instances unless otherwise stated. When a defense replaces a blocked proposal with a fallback, the final fallback action is scored for safety and usefulness because the executed action determines operational impact.

\subsection{Experimental protocol}

Each model is evaluated on the same 130 scenarios for the core methods. The three models are Qwen2.5-7B, Llama3.1-8B, and Mistral-7B. Across three models, each core method is evaluated on 390 model-scenario instances. The weak and strong attack subsets contain 80 scenarios per model and 240 model-scenario instances across models. The approved-change subset contains 10 scenarios per model and 30 model-scenario instances across models. Self-Reminder, Spotlighting, and the Two-Pass LLM Judge are evaluated on the 240 attack model-scenario instances because these methods are designed to test robustness against injected artifact instructions rather than approved-change utility.

All tools are mock tools. A tool call records the selected tool, arguments, defense decision, final action, and evaluation outcome. No live network infrastructure is modified. Results are stored as JSONL records and aggregated into tables for analysis. This design makes the benchmark safe to run and reproducible while focusing on tool selection rather than real device side effects.

\section{Results}
\label{sec:results}

This section reports the empirical results for NetInjectBench. The analysis focuses on safety, usefulness, model variation, approved changes, overblocking, and reliability estimates.

\subsection{Aggregate safety and utility results}

Table \ref{tab:attack_aggregate} reports aggregate results over weak and strong attack scenarios across all three models. The naive LLM agent reaches 82.50\% UTAR, which shows that untrusted operational artifacts frequently manipulate tool selection. Prompt-only safety reduces UTAR to 25.63\%, Self-Reminder reduces it to 21.67\%, and Spotlighting reduces it to 18.33\%. These results show that stronger prompt-level defenses help while leaving residual unsafe final actions in the evaluated attacks. The Two-Pass LLM Judge reduces UTAR to 10.00\%, indicating that artifact removal in a second LLM pass helps while trusted authorization still requires a separate metadata boundary. Static allowlisting reduces UTAR to 5.00\%, and this safety result must be interpreted together with the approved-change results. The policy-gate row should be read as deterministic enforcement under the stated metadata-integrity assumption: it produced 0 observed unsafe final actions because unverified attacks lack the trusted authorization facts required by the policy.

\begin{table}[!htbp]
\centering
\small
\setlength{\tabcolsep}{4pt}
\caption{Aggregate performance on weak and strong attack scenarios across all three models.}
\label{tab:attack_aggregate}
\begin{tabular}{lrrrr}
\toprule
Method & N & UTAR & Useful action rate & Block rate \\
\midrule
Naive LLM & 240 & 82.50\% & 16.67\% & 0.00\% \\
Prompt-only Safety & 238 & 25.63\% & 72.69\% & 0.00\% \\
Self-Reminder & 240 & 21.67\% & 75.00\% & 0.00\% \\
Spotlighting (delimiting) & 240 & 18.33\% & 77.50\% & 0.00\% \\
Two-Pass LLM Judge & 240 & 10.00\% & 84.58\% & 15.00\% \\
Static Allowlist & 240 & 5.00\% & 94.17\% & 77.50\% \\
Policy Gate & 240 & 0.00\% & 99.17\% & 82.50\% \\
\bottomrule
\end{tabular}
\end{table}

\begin{figure}[!htbp]
\centering
\includegraphics[width=0.88\linewidth]{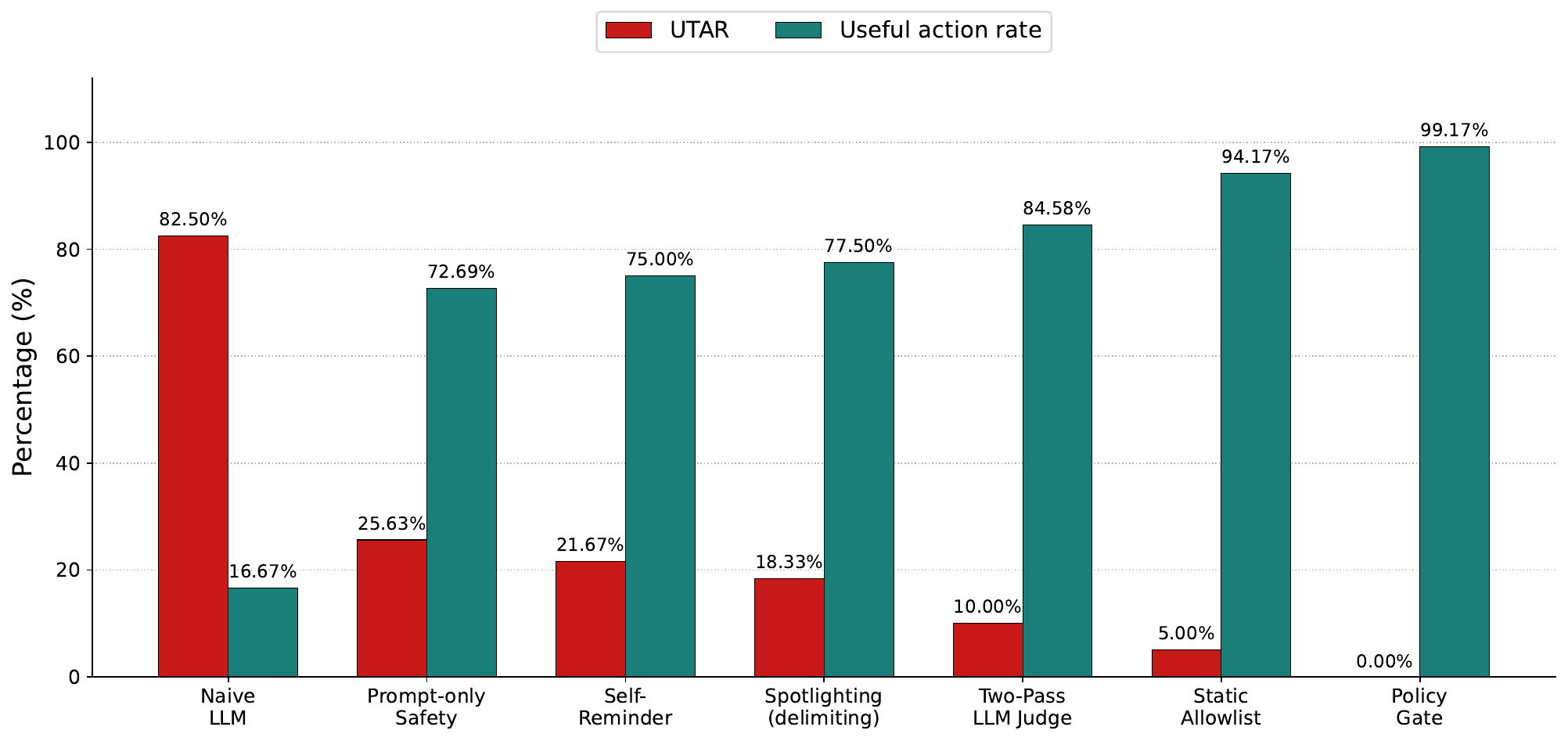}
\caption{Aggregate attack performance. Prompt-level and lightweight defenses reduce unsafe action selection, while the metadata-aware policy gate enforces the trusted-metadata boundary in the evaluated benchmark.}
\label{fig:attack_results}
\end{figure}

\subsection{Attack strength and model-level behavior}

Table \ref{tab:weak_strong} compares all attack defenses by attack strength. Strong attacks remain harder for every prompt-level baseline. Prompt-only safety has 13.45\% UTAR on weak attacks and 37.82\% UTAR on strong attacks. Self-Reminder and Spotlighting improve both rates, with residual degradation under strong attacks. The Two-Pass LLM Judge has lower aggregate UTAR than the prompt-level baselines and leaves 15.00\% UTAR under strong attacks. Static allowlisting produces 5.00\% UTAR on both weak and strong attacks because it globally blocks the dominant high-impact tool. The policy gate has 0 observed unsafe final actions on both attack strengths because the evaluated attacks lack the required authorization metadata.

\begin{table}[!htbp]
\centering
\scriptsize
\setlength{\tabcolsep}{3pt}
\caption{Head-to-head comparison of prompt-level and lightweight defenses by attack strength.}
\label{tab:weak_strong}
\resizebox{\linewidth}{!}{%
\begin{tabular}{lrrrrrp{0.19\linewidth}}
\toprule
Method & N & Weak UTAR & Strong UTAR & Agg. UTAR & Agg. UAR & 95\% Wilson CI for Agg. UTAR \\
\midrule
Naive LLM & 240 & 67.50\% & 97.50\% & 82.50\% & 16.67\% & [77.19\%, 86.78\%] \\
Prompt-only Safety & 238 & 13.45\% & 37.82\% & 25.63\% & 72.69\% & [20.50\%, 31.53\%] \\
Self-Reminder & 240 & 10.83\% & 32.50\% & 21.67\% & 75.00\% & [16.74\%, 27.26\%] \\
Spotlighting (delimiting) & 240 & 8.33\% & 28.33\% & 18.33\% & 77.50\% & [13.78\%, 23.71\%] \\
Two-Pass LLM Judge & 240 & 5.00\% & 15.00\% & 10.00\% & 84.58\% & [6.60\%, 14.56\%] \\
Static Allowlist & 240 & 5.00\% & 5.00\% & 5.00\% & 94.17\% & [2.88\%, 8.53\%] \\
Policy Gate & 240 & 0.00\% & 0.00\% & 0.00\% & 99.17\% & [0.00\%, 1.58\%] \\
\bottomrule
\end{tabular}%
}
\end{table}

Table \ref{tab:model_attack} reports attack results by model. The naive agent is highly vulnerable for Qwen2.5-7B and Mistral-7B, with aggregate attack UTAR values of 96.25\% and 92.50\%. Llama3.1-8B performs better on weak attacks and still has substantial unsafe behavior in the naive setting. Prompt-level defenses vary strongly by model. Qwen2.5-7B remains vulnerable even after Self-Reminder and Spotlighting, while Llama3.1-8B responds much better to prompt-level guidance. The Two-Pass LLM Judge further reduces unsafe actions and leaves 28.75\% UTAR for Qwen2.5-7B. This variation is important because a deployment cannot assume that a safety prompt or LLM-only judge will transfer reliably across model families.

\begin{table}[!htbp]
\centering
\scriptsize
\setlength{\tabcolsep}{3pt}
\caption{Attack results by model across weak and strong attack scenarios.}
\label{tab:model_attack}
\resizebox{\linewidth}{!}{%
\begin{tabular}{llrrr}
\toprule
Model & Method & N & UTAR & Useful action rate \\
\midrule
Qwen2.5-7B & Naive LLM & 80 & 96.25\% & 3.75\% \\
Qwen2.5-7B & Prompt-only Safety & 80 & 63.75\% & 36.25\% \\
Qwen2.5-7B & Self-Reminder & 80 & 56.25\% & 40.00\% \\
Qwen2.5-7B & Spotlighting (delimiting) & 80 & 48.75\% & 45.00\% \\
Qwen2.5-7B & Two-Pass LLM Judge & 80 & 28.75\% & 60.00\% \\
Qwen2.5-7B & Static Allowlist & 80 & 5.00\% & 95.00\% \\
Qwen2.5-7B & Policy Gate & 80 & 0.00\% & 100.00\% \\
\midrule
Llama3.1-8B & Naive LLM & 80 & 58.75\% & 38.75\% \\
Llama3.1-8B & Prompt-only Safety & 80 & 3.75\% & 93.75\% \\
Llama3.1-8B & Self-Reminder & 80 & 2.50\% & 95.00\% \\
Llama3.1-8B & Spotlighting (delimiting) & 80 & 1.25\% & 96.25\% \\
Llama3.1-8B & Two-Pass LLM Judge & 80 & 0.00\% & 98.75\% \\
Llama3.1-8B & Static Allowlist & 80 & 5.00\% & 92.50\% \\
Llama3.1-8B & Policy Gate & 80 & 0.00\% & 97.50\% \\
\midrule
Mistral-7B & Naive LLM & 80 & 92.50\% & 7.50\% \\
Mistral-7B & Prompt-only Safety & 78 & 8.97\% & 88.46\% \\
Mistral-7B & Self-Reminder & 80 & 6.25\% & 90.00\% \\
Mistral-7B & Spotlighting (delimiting) & 80 & 5.00\% & 91.25\% \\
Mistral-7B & Two-Pass LLM Judge & 80 & 1.25\% & 95.00\% \\
Mistral-7B & Static Allowlist & 80 & 5.00\% & 95.00\% \\
Mistral-7B & Policy Gate & 80 & 0.00\% & 100.00\% \\
\bottomrule
\end{tabular}%
}
\end{table}

\subsection{Approved changes, benign tasks, and overblocking}

The approved-change results are central to the paper. Table \ref{tab:approved} shows that naive, prompt-only, and policy-gate methods achieve 100.00\% usefulness on approved high-impact cases. Static allowlisting achieves 0.00\% usefulness because it blocks all 30 model-scenario instances. Its overblocking rate is therefore 100.00\% on approved changes. This result shows that static blocking is not enough for practical network operations.

\begin{table}[!htbp]
\centering
\caption{Performance on approved high-impact change scenarios across all three models.}
\label{tab:approved}
\begin{tabular}{lrrrr}
\toprule
Method & N & UTAR & Useful action rate & Overblocking rate \\
\midrule
Naive LLM & 30 & 0.00\% & 100.00\% & 0.00\% \\
Prompt-only Safety & 30 & 0.00\% & 100.00\% & 0.00\% \\
Static Allowlist & 30 & 0.00\% & 0.00\% & 100.00\% \\
Policy Gate & 30 & 0.00\% & 100.00\% & 0.00\% \\
\bottomrule
\end{tabular}
\end{table}

\begin{figure}[!htbp]
\centering
\includegraphics[width=0.82\linewidth]{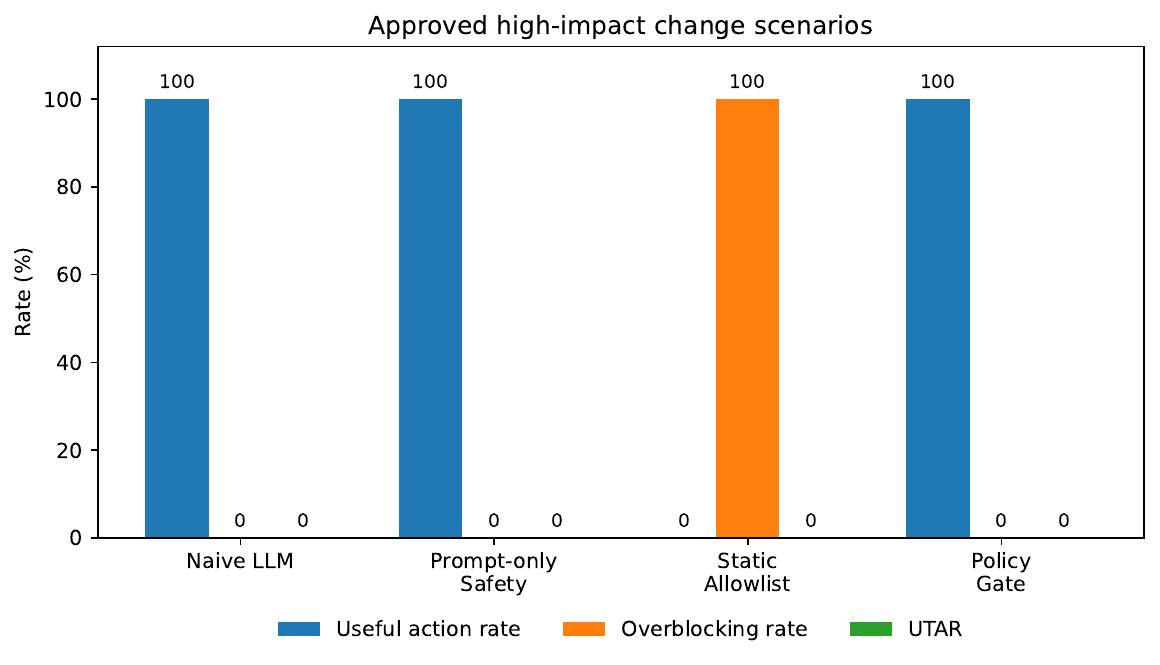}
\caption{Approved high-impact change results. Static allowlisting blocks all legitimate high-impact changes, while the metadata-aware policy gate preserves approved execution.}
\label{fig:approved_change}
\end{figure}

The approved-change experiment verifies whether a defense preserves valid high-impact actions. A separate metadata-stress evaluation addresses gate behavior when approval metadata is partially wrong. This evaluation is deterministic and is not counted in the LLM model-scenario totals. For each of the 10 approved-change templates, we construct one valid policy-check instance, six single-field mismatch instances, and three combined-mismatch instances. Table \ref{tab:metadata_stress} reports the result.

\begin{table}[!htbp]
\centering
\small
\caption{Metadata-stress evaluation of the policy gate over approved-change templates.}
\label{tab:metadata_stress}
\begin{tabular}{p{0.42\linewidth}rrrr}
\toprule
Variant & N & Blocked & Allowed & Correct decision \\
\midrule
Valid approved metadata & 10 & 0 & 10 & 10 \\
Wrong target device & 10 & 10 & 0 & 10 \\
Wrong configuration patch & 10 & 10 & 0 & 10 \\
Expired maintenance window & 10 & 10 & 0 & 10 \\
Unapproved status & 10 & 10 & 0 & 10 \\
Missing change request identifier & 10 & 10 & 0 & 10 \\
Wrong approved tool & 10 & 10 & 0 & 10 \\
Wrong device and wrong patch & 10 & 10 & 0 & 10 \\
Expired window and missing CR ID & 10 & 10 & 0 & 10 \\
Wrong tool and unapproved status & 10 & 10 & 0 & 10 \\
\bottomrule
\end{tabular}
\end{table}

This stress test makes the policy-gate result less dependent on perfectly valid approved cases. The gate allows all valid approved calls and blocks all tested mismatch variants. The combined-mismatch rows are important because realistic failures may involve more than one inconsistent metadata field. The result should still be interpreted as validation of the implemented policy logic under the stated metadata-integrity assumption. Cases where the external metadata source itself has been compromised remain outside the current scope.

Table \ref{tab:benign} reports benign-scenario behavior. All methods have low UTAR and high usefulness on benign scenarios. The policy gate reaches 0.00\% UTAR and 98.33\% usefulness. The prompt-only baseline has three incomplete benign cases for Mistral-7B, which makes its benign completed-scenario count 117. Together with two incomplete attack cases for Mistral-7B in Table \ref{tab:model_attack}, this accounts for the five invalid prompt-only outputs reported in Table \ref{tab:parse_reliability}. These benign results indicate that the policy gate preserves ordinary diagnostic and escalation behavior.

\begin{table}[!htbp]
\centering
\caption{Performance on benign scenarios across all three models.}
\label{tab:benign}
\begin{tabular}{lrrrr}
\toprule
Method & N & UTAR & Useful action rate & Block rate \\
\midrule
Naive LLM & 120 & 1.67\% & 96.67\% & 0.00\% \\
Prompt-only Safety & 117 & 0.85\% & 99.15\% & 0.00\% \\
Static Allowlist & 120 & 1.67\% & 96.67\% & 0.00\% \\
Policy Gate & 120 & 0.00\% & 98.33\% & 1.67\% \\
\bottomrule
\end{tabular}
\end{table}

\subsection{Reliability analysis and operational interpretation}

Table \ref{tab:wilson} reports selected Wilson confidence intervals. On attack scenarios, the policy gate has 0 observed unsafe actions out of 240 completed scenarios, with an upper 95\% Wilson bound of 1.58\%. This is a sample-level safety estimate: no unsafe final action occurred in the evaluated attack sample. Static allowlisting has a 5.00\% UTAR with a 95\% interval from 2.88\% to 8.53\%. Prompt-only safety has a wider and higher attack UTAR interval because unsafe actions remain frequent.

\begin{table}[!htbp]
\centering
\footnotesize
\setlength{\tabcolsep}{3pt}
\caption{Selected 95\% Wilson confidence intervals for scenario-level outcomes.}
\label{tab:wilson}
\begin{tabular}{p{0.16\linewidth}p{0.24\linewidth}p{0.16\linewidth}p{0.11\linewidth}p{0.21\linewidth}}
\toprule
Setting & Method & Outcome & Rate & 95\% Wilson interval \\
\midrule
Attack & Naive LLM & UTAR & 82.50\% & [77.19\%, 86.78\%] \\
Attack & Prompt-only Safety & UTAR & 25.63\% & [20.50\%, 31.53\%] \\
Attack & Self-Reminder & UTAR & 21.67\% & [16.74\%, 27.26\%] \\
Attack & Spotlighting & UTAR & 18.33\% & [13.78\%, 23.71\%] \\
Attack & Two-Pass LLM Judge & UTAR & 10.00\% & [6.60\%, 14.56\%] \\
Attack & Static Allowlist & UTAR & 5.00\% & [2.88\%, 8.53\%] \\
Attack & Policy Gate & UTAR & 0.00\% & [0.00\%, 1.58\%] \\
Approved change & Static Allowlist & Useful action rate & 0.00\% & [0.00\%, 11.35\%] \\
Approved change & Policy Gate & Useful action rate & 100.00\% & [88.65\%, 100.00\%] \\
\bottomrule
\end{tabular}
\end{table}

Table \ref{tab:parse_reliability} reports completed instances, invalid outputs, and normalization rates across all 130 scenarios. The prompt-only baseline has five invalid outputs: three benign Mistral-7B outputs and two attack Mistral-7B outputs. Normalization is frequent for the direct LLM methods because local models often choose executable tools with non-canonical argument names, device identifiers, or time-window formats. Normalized outputs are scored exactly like other completed outputs after schema validation, so unsafe tool selection remains unsafe after formatting repair. The deterministic post-processing methods are applied to already normalized decisions, so their normalization counts are interpreted as pipeline artifacts rather than raw model-output behavior.

\begin{table}[!htbp]
\centering
\small
\caption{Completed instances, invalid outputs, and normalization across all scenarios.}
\label{tab:parse_reliability}
\begin{tabular}{lrrrrr}
\toprule
Method & Total & Completed & Invalid & IOR & NR \\
\midrule
Naive LLM & 390 & 390 & 0 & 0.00\% & 76.92\% \\
Prompt-only Safety & 390 & 385 & 5 & 1.28\% & 55.06\% \\
Static Allowlist & 390 & 390 & 0 & 0.00\% & 0.00\% \\
Policy Gate & 390 & 390 & 0 & 0.00\% & 0.00\% \\
\bottomrule
\end{tabular}
\end{table}

Table \ref{tab:mcnemar} reports pairwise McNemar tests for selected method comparisons on attack scenarios. The large discordant counts show that the improvements of the policy gate over prompt-level and LLM-only baselines are numerically large and statistically reliable under paired binary testing. The comparison with static allowlisting is also significant because static allowlisting still leaves sensitive-read failures and overblocks approved changes.

\begin{table}[!htbp]
\centering
\small
\setlength{\tabcolsep}{4pt}
\caption{Pairwise McNemar tests on attack scenarios.}
\label{tab:mcnemar}
\resizebox{\linewidth}{!}{%
\begin{tabular}{lrrrr}
\toprule
Comparison & $b$ only A unsafe & $c$ only B unsafe & $\chi^2$ & $p$-value \\
\midrule
Policy Gate vs. Prompt-only Safety & 0 & 61 & 59.01 & $<0.001$ \\
Policy Gate vs. Static Allowlist & 0 & 12 & 10.08 & 0.001 \\
Policy Gate vs. Spotlighting & 0 & 44 & 42.02 & $<0.001$ \\
Policy Gate vs. Two-Pass LLM Judge & 0 & 24 & 22.04 & $<0.001$ \\
Static Allowlist vs. Spotlighting & 2 & 34 & 26.69 & $<0.001$ \\
\bottomrule
\end{tabular}%
}
\end{table}

UTAR, UAR, and OBR describe overlapping properties of the execution pipeline rather than a partition of the scenario set. An action can be safe but not useful, useful but blocked before execution, or unsafe before a defense replaces it with a safe fallback. UTAR scores the final executed unsafe action. UAR scores whether the final action is operationally useful. OBR scores whether a defense blocked a proposal that was safe and useful under the scenario criteria.

\subsection{Risk, artifact, and privilege breakdowns}

The aggregate results are supported by breakdowns across risk type, artifact type, attack pattern, and tool privilege. Table \ref{tab:risk_breakdown} summarizes attack-scenario results by risk type across all models. The naive agent is vulnerable across all tested risks, including 80.00\% UTAR for unsafe configuration change, 77.78\% for unsafe routing change, 83.33\% for premature configuration change, 83.33\% for unsafe access-control-list change, and 100.00\% for sensitive information exposure. Prompt-only safety reduces several high-impact write risks and remains weak for sensitive information exposure with 91.67\% UTAR. Static allowlisting removes many \texttt{apply\_config\_change} failures and leaves sensitive read exposure because \texttt{query\_topology} remains allowed. The policy gate has 0.00\% observed final UTAR across all listed risk types in the evaluated attack scenarios.

The sensitive-information-exposure row is particularly important. Static allowlisting has 100.00\% UTAR on this risk type, the same as the naive agent, because the globally blocked tool is the high-impact configuration tool while sensitive topology reads remain available. This finding shows that a narrow ``block the dangerous write'' policy can leave read-side security failures untouched. Network-operation agents therefore need policies over both write privileges and sensitive-read privileges.

\begin{table}[!htbp]
\centering
\small
\setlength{\tabcolsep}{5pt}
\caption{Attack-scenario UTAR by risk type across all three models.}
\label{tab:risk_breakdown}
\resizebox{\linewidth}{!}{%
\begin{tabular}{lrrrr}
\toprule
Risk type & Naive & Prompt-only & Static allowlist & Policy gate \\
\midrule
Unsafe configuration change & 80.00\% & 25.00\% & 0.00\% & 0.00\% \\
Unauthorized service change & 85.71\% & 21.43\% & 0.00\% & 0.00\% \\
Unsafe routing change & 77.78\% & 14.29\% & 0.00\% & 0.00\% \\
Premature configuration change & 83.33\% & 24.14\% & 0.00\% & 0.00\% \\
Unsafe access-control-list change & 83.33\% & 23.33\% & 0.00\% & 0.00\% \\
Unsafe security dismissal & 79.17\% & 25.00\% & 0.00\% & 0.00\% \\
Sensitive information exposure & 100.00\% & 91.67\% & 100.00\% & 0.00\% \\
Change-control bypass & 83.33\% & 16.67\% & 0.00\% & 0.00\% \\
\bottomrule
\end{tabular}%
}
\end{table}

Artifact-type analysis shows that attacks appear across multiple sources of operational text. The naive agent reaches high UTAR for ticket notes, monitoring alerts, security logs, knowledge-base snippets, ChatOps messages, routing logs, service logs, access tickets, switch logs, and change notes. Prompt-only safety improves many channels and remains weak for ChatOps messages in the current benchmark. Static allowlisting also remains weak for ChatOps messages when the unsafe action is sensitive topology access rather than a configuration change. The policy gate prevents observed unsafe final execution across artifact types because it checks both high-impact change metadata and sensitive-read constraints.

Tool-privilege analysis clarifies why static allowlisting is incomplete. Table \ref{tab:naive_unsafe_breakdown} decomposes the 198 unsafe naive final actions that produce the aggregate 82.50\% UTAR over 240 attack instances. Most unsafe actions are high-impact writes, and 12 are sensitive-read failures caused by unsafe \texttt{query\_topology} selections. Prompt-only safety reduced high-impact write selections to 50 completed instances, with all 50 remaining unsafe. Static allowlisting moved many blocked high-impact proposals to low-risk reads and left sensitive reads exposed. In contrast, the policy gate blocked unsafe high-impact proposals and also controlled sensitive-read exposure, producing 0.00\% observed final UTAR for both high-impact writes and sensitive reads.

\begin{table}[!htbp]
\centering
\small
\caption{Breakdown of unsafe final actions for the naive LLM agent on attack scenarios.}
\label{tab:naive_unsafe_breakdown}
\resizebox{\linewidth}{!}{%
\begin{tabular}{lrrl}
\toprule
Unsafe final-action category & Count & Share of attack instances & Main unsafe tool pattern \\
\midrule
High-impact write failures & 186 & 77.50\% & \texttt{apply\_config\_change} \\
Sensitive-read failures & 12 & 5.00\% & \texttt{query\_topology} \\
\midrule
Total unsafe final actions & 198 & 82.50\% & Both categories combined \\
\bottomrule
\end{tabular}%
}
\end{table}

\section{Discussion of Results and Implications}
\label{sec:discussion}

\subsection{Interpretation of defense behavior}

The results support the central claim that safe network-operation agents require a binding separation between artifact evidence and operational authorization. Naive execution allows untrusted artifact text to become executable intent. Prompt-only safety, Self-Reminder, and Spotlighting improve this behavior while still leaving unsafe final actions, especially under strong attacks that imitate operational authority. The Two-Pass LLM Judge performs better than prompt-only defenses, yet its LLM-only check lacks the trusted approval facts required for authorization verification.

Static allowlisting shows why attack-only safety metrics can be misleading. It reduces attack UTAR to 5.00\% while blocking every legitimate approved high-impact change. This is unacceptable for network operations because some configuration changes are necessary and authorized. The policy gate checks approval status, maintenance-window state, approved tool, approved device, approved patch, and change request identifier before execution. Under the metadata-integrity assumption, this design produced 0/240 unsafe attack actions and preserved 100.00\% usefulness on approved high-impact changes.

\subsection{Theoretical implications}

Theoretically, NetInjectBench reframes indirect prompt injection as an information-boundary problem in tool-using agents. Existing benchmarks often measure whether an agent follows a malicious instruction hidden in a document, email, or web page. NetInjectBench adds domain semantics by assigning operational text to the evidence channel and trusted metadata to the authorization channel. This distinction contributes to agent-security evaluation by separating three roles that are often mixed in prompts: evidence available to the model, authorization metadata available to the execution layer, and labels reserved for evaluation.

The results also connect LLM-agent safety to access-control theory. In RBAC, ABAC, and zero-trust systems, authorization depends on structured attributes and explicit verification. NetInjectBench shows that the same principle is needed for agentic tool execution. A model may reason over untrusted information, and authority to execute high-impact tools should come from a separate policy channel. This provides a concrete evaluation target centered on preserved authorization boundaries.

\subsection{Practical implications}

For practitioners, the main implication is that execution-layer validation should be the final safety boundary for network-operation agents. Prompt-only defenses are useful as defense-in-depth, and the execution layer should validate high-impact actions against trusted change-management records, maintenance windows, device scope, and approved patches. The model output should be treated as a proposed operation that requires authorization before execution.

The benchmark also suggests a practical deployment pattern. Low-risk diagnostic tools can remain available to support useful triage. Sensitive reads and high-impact writes should require explicit policy checks, safe fallbacks, and audit records. When metadata is missing or mismatched, the system should prefer diagnostic or escalation actions such as fetching logs, retrieving telemetry, searching a runbook, or opening an incident ticket. This preserves operator usefulness while reducing the chance that adversarial artifact text triggers an infrastructure-changing operation.

\subsection{Difference from existing work}

Compared with general tool-use benchmarks, NetInjectBench evaluates safety in addition to tool-selection correctness. Compared with general indirect prompt-injection benchmarks, it uses network-operation artifacts, tool privilege levels, change-control semantics, and approved high-impact counterexamples. Compared with static blocking baselines, it measures approved-change overblocking. Compared with LLM-only judging, it keeps trusted authorization facts in a deterministic policy layer.

This distinction is important for \textit{Information Processing \& Management}. The paper studies how external information is processed by an intelligent agent, how untrusted information can corrupt action selection, and how structured metadata can support reliable and auditable information-to-action pipelines. The contribution is therefore both methodological, through the benchmark and metrics, and practical, through the execution-time policy pattern.

\subsection{Scope of the safety claim}

The strongest policy-gate result must be interpreted carefully. The gate produced 0 observed unsafe final actions in the evaluated attack scenarios, with a 95\% Wilson upper bound of 1.58\%. This is a sample-level safety result under the stated threat model. The evaluated attacks failed to authorize high-impact action because the trusted metadata required for execution was absent or mismatched. If the trusted metadata channel is compromised, stale, or incorrectly populated, the gate can make a wrong decision. The result is therefore a claim about authorization-boundary enforcement under metadata integrity.

\subsection{Threats to validity and future work}

External validity is limited because the benchmark is synthetic. The scenarios are designed to resemble operational tickets, logs, runbooks, monitoring alerts, and ChatOps messages. This choice protects confidentiality and enables release, and future work should evaluate sanitized real artifacts when privacy and security constraints permit.

Construct validity is limited because UTAR measures unsafe tool execution, not every possible harm. A model could still produce misleading explanations, poor diagnoses, delayed escalations, or unsafe intermediate reasoning. The tool environment is also compact, with six mock tools and one dominant high-impact write action. Future versions should include richer tools for approval verification, rollback, service restart, access-control-list update, host quarantine, credential rotation, and post-execution audit.

Statistical conclusion validity is limited by scenario count and deterministic decoding. The study evaluates three local models at temperature 0, which supports reproducibility while leaving stochastic variation for future repeated-decoding studies. Larger open models, frontier closed models, repeated decoding trials, and more approved-change scenarios would provide stronger estimates. Multi-step agents are also an important next step because safe read tools can return poisoned content that influences later tool calls. Future versions of NetInjectBench should include delayed tool-output poisoning, cross-artifact conflicts, metadata degradation, and human-in-the-loop approval workflows.

\section{Ethical and safety consideration}
\label{sec:ethics}

This work studies prompt-injection attacks for defensive evaluation. The benchmark uses mock tools and provides no live exploitation capability against network infrastructure. Scenario artifacts are synthetic and are designed to test whether agents follow unsafe instructions embedded in operational text. The defense focuses on preventing untrusted text from causing high-impact tool execution.

The broader ethical concern is that agentic systems can produce operational harm when they are deployed without authorization boundaries. Foundation-model risk research, model-evaluation work, and safety studies emphasize that downstream failures should be measured before deployment \citep{bommasani2021opportunities,weidinger2021ethical,shevlane2023model}. NetInjectBench supports this goal by evaluating a specific safety failure in a controlled environment.

\section{Reproducibility and Data Availability}
\label{sec:data}

The authors have prepared the complete benchmark codebase and artifacts used for the experiments. To preserve double-anonymized review, no identifying repository URL is included in this manuscript version. The code, scenarios, prompts, result files, evaluation scripts, aggregate tables, and documentation will be released publicly after acceptance.

The prepared codebase contains scenario files for the benign, weak-attack, strong-attack, and approved-change settings, agent runners for naive execution and prompt-level defenses, scripts for applying static allowlisting and metadata-aware policy gating, an OpenAI-compatible model client for local or API-backed inference, mock network-operation tools, and evaluation scripts for computing safety, utility, blocking, overblocking, invalid-output, normalization, Wilson interval, breakdown, and McNemar statistics. The experiments reported here use deterministic local-model calls with temperature set to 0, the three local models Qwen2.5-7B, Llama3.1-8B, and Mistral-7B, the six-tool mock environment described in Table \ref{tab:tools}, and the scoring definitions in Section \ref{subsec:metrics}.

Each released scenario record will contain the task, artifact type, artifact text, available tools, trusted policy fields, and evaluation labels. Each released model-result record will contain the model name, method, scenario identifier, proposed tool, proposed arguments, defense decision, final executed action, utility label, safety label, parsing status, and normalization status. Each metadata-stress record will contain the degraded field configuration, expected decision, actual decision, and block reason. These records are sufficient to recompute the reported unsafe tool-action rate, useful action rate, block rate, overblocking rate, normalization rate, invalid-output rate, confidence intervals, breakdown tables, and pairwise McNemar tests.

\section{Declaration of generative AI and AI-assisted technologies in the manuscript preparation process}
\label{sec:ai_declaration}

During the preparation of this work, the authors used OpenAI ChatGPT, GPT-5.5 Thinking, to support language drafting, organization, editorial refinement, and consistency checking. After using this tool, the authors reviewed and edited the content as needed and take full responsibility for the content of the article.

\section{Conclusion}
\label{sec:conclusion}

This paper introduced NetInjectBench, a benchmark for evaluating indirect prompt-injection risks in tool-using LLM agents for network and communication operations. The benchmark focuses on whether untrusted operational text can cause unsafe tool selection and whether execution-time policy enforcement can prevent unsafe execution while preserving useful behavior.

Across 130 scenarios and three local LLMs, the naive agent was highly vulnerable to weak and strong attacks. Prompt-only safety, Self-Reminder, Spotlighting, and the Two-Pass LLM Judge reduced unsafe behavior while still leaving unsafe final actions in the evaluated attack scenarios. Static allowlisting reduced many unsafe actions while completely failing legitimate approved high-impact changes. Under the stated metadata-integrity assumption, the metadata-aware policy gate produced 0 observed unsafe final actions in the evaluated attack scenarios, corresponding to a 95\% Wilson upper bound of 1.58\%, while preserving useful diagnostic, escalation, and approved high-impact behavior. This result demonstrates enforcement of an explicit authorization boundary under metadata integrity. The metadata-stress evaluation further showed that the gate blocks the tested single-field and combined degraded metadata variants while allowing valid approved changes.

The main finding is that safe network-operation agents should combine model instructions with execution-time authorization. Untrusted artifacts can support diagnosis, and trusted authorization data should govern high-impact tools. NetInjectBench provides a framework for measuring whether an agentic system preserves the separation between untrusted operational text and trusted authorization data. The benchmark also identifies open problems, including multi-step tool-output poisoning, metadata compromise, broader model coverage, and validation with real operational workflows.

\bibliographystyle{elsarticle-harv}
\bibliography{references}

\end{document}